\documentclass[preprint,aps,superscriptaddress,nofootinbib]{revtex4}
\usepackage{amsmath}
\usepackage{graphicx}
\usepackage{color}
\usepackage{slashed}

\begin{document}

\title{Nonthermal correction to black hole spectroscopy}

\author{Wen-Yu Wen}\thanks{%
E-mail: steve.wen@gmail.com}
\affiliation{Department of Physics and Center for High Energy Physics, Chung Yuan Christian University, Chung Li City, Taiwan}
\affiliation{Leung Center for Cosmology and Particle Astrophysics\\
National Taiwan University, Taipei 106, Taiwan}

\begin{abstract}
Area spectrum of black holes have been obtained via various methods such as quasinormal modes, adiabatic invariance and angular momentum.  Among those methods, calculations were done by assuming black holes in thermal equilibirum.   Nevertheless, black holes in the asymptotically flat space usually have negative specific heat and therefore tend to stay away from thermal equilibrium.  Even for those black holes with positive specific heat, temperature may still not be well defined in the process of radiation, due to the back reaction of decreasing mass.  Respect to these facts, it is very likely that Hawking radiation is nonthermal and the area spectrum is no longer equidistant.  In this note, we would like to illustrate how the area spectrum of black holes is corrected by this nonthermal effect.
\end{abstract}

\maketitle



A finite size system often displays a discrete energy spectrum for quantum fluctuation.   It was suggested that since the dynamics of a black hole is uniquely determined by its charge(s), which is closely related to the finite region enclosed by the horizon, one also expects the mass or area spectrum to display similar discreteness\cite{Bekenstein:1995ju,Bekenstein:1997bt}.  There were many proposals to obtain area spectrum for various black holes since then.  Most earlier methods of quantizing horizon area are based on real or imaginary part of quasinormal modes\cite{Hod:1998vk,Hod:2003jn,Barvinsky:2001tw,Maggiore:2007nq,Vagenas:2008yi,
LopezOrtega:2010tg,Setare:2003bd,Setare:2004uu}.   Recently the application of adiabatic invariant action variable did not use the quasinormal modes\cite{Majhi:2011gz,Zeng:2012wb} and the idea of quantizing angular momentum to obtain area spectrum first appeared in the study of non-extremal RN black holes\cite{Ropotenko:2009mh}.  The various methods of quantization have settled down to a spectrum of equidistant discreteness
\begin{equation}
\Delta A = c \l_p^2.
\end{equation}
In particular, one obtained $c = 8\pi$ for various kinds of black holes in different spacetime dimensions.  Nevertheless, this universal result is closely related to the assumption that the black hole is in the thermal equilibrium state where the Hawking temperature is well defined.  Realistic black holes are more likely to be in the nonequilibrium state due to its negative specific heat.  Even for those black holes with positive specific heat, temperature may still be ill-defined in the process of radiation, due to the back reaction of decreasing mass.  A universal  logarithm correction to the Bekenstein-Hawking area law has been predicted in various theories of quantum gravity and modified general relativity, such that
\begin{equation}\label{log_correction}
S_{BH} = \frac{A}{4 \l_p^2} + \alpha \ln (\frac{A}{\l_p^2}).
\end{equation}
The corresponded correction to the area spectrum was computed for $\alpha = -\frac{3}{2}$ in the context of adiabatic invaraince approach for a constant surface gravity and uneven discreteness was observed\cite{Majhi:2011gz}:
\begin{equation}
\Delta A \simeq 8\pi \l_p^2 - \frac{32 \pi\alpha \l_p^4}{A}.
\end{equation}
While the above logarithmic correction in (\ref{log_correction}) can be regarded as the consequence of loop quantum correction of surface gravity\cite{Work:1985,Lousto:1988}, we are looking for the other correction due to back reaction from the Hawking radiation.   Among various models of black hole radiation, the tunneling model proposed by Parikh and Wilczek \cite{Parikh:1999mf} has provided useful insights in the effort to resolve the Information Loss Paradox \cite{Zhang:2009jn}, black hole evolution  \cite{Kim:2013qxu,Chatrabhuti:2014fpa} and black hole remnant \cite{Chen:2009ut}.  The Parikh-Wilczek model regards the Hawking radiation as a tunneling process in some stationary vacuum.  The potential barrier is dynamically establsihed due to the back reaction which observes energy conservation.  The emission rate in the tunneling model has a universal result
\begin{equation}\label{tunnel_rate}
\Gamma \sim e^{\Delta S_{BH}},
\end{equation} 
given the black hole entropy change $\Delta S_{BH}$ due to radiation.  The back reaction constantly changes the surface gravity during tunneling process, therefore the black hole is never in the thermal equilibrium.  In the following, we would like to use the Schwarzschild black hole as an example to argue that the back reaction effect could produce another correction to area spectrum of order ${\cal O}(A^{-1})$.  

In the case of Schwarzschild black hole of mass $M$, we have the change of entropy
\begin{equation}
\hbar\Delta S \simeq -8\pi M\omega + 4\pi \omega^2 - \alpha \hbar(2 \frac{\omega}{M}+\frac{\omega^2}{M^2} ) + {\cal O}(\alpha^2).
\end{equation}  
where the first term is nothing but the thermal spectrum if the inverse of Hawking temperature $T_H^{-1}=8\pi M$ is identified, and we regard the second term as the nonthermal correction due to back reaction.  Those terms with $\alpha$ inside are the series expansion of the logarithmic correction with respect to large black hole mass and we reagrd them as the quantum correction to the spectrum. 
To see how the area spectrum also receives correction from those nonthermal and quantum effect, let us recall the derivation of (\ref{tunnel_rate}) and then divert to the quantization of area.   The tunneling process happens at the horizon in the following metric\cite{Parikh:1999mf} in the unit $G=c=k_B=1:$\footnote{We remark that one cannot further set $\hbar=1$ as in the nature unit.  Since in the unit $G=c=1$, one has dimension $[L]=[T]=[M]$, and $\hbar$ has dimension $[L]^2$.  We are grateful to Otto Kong to point out this.}
\begin{equation}
ds^2 = -(1-\frac{2M}{r})dt^2 + 2\sqrt{\frac{2M}{r}}dtdr +dr^2 + r^2 d\Omega_2^2,
\end{equation}
which can be obtained from the static Schwarzschild black hole metric by a coordinate transformation of the Painlev\`{e}-type.  The WKB approximation states that the emission rate $\Gamma \sim e^{-2 Im S/\hbar}$, where the imaginary part of action reads
\begin{equation}\label{integral}
Im S = Im \int_{r_{in}}^{r_{out}}{p_r dr} = Im \int_M^{M-\omega}\int_{2M}^{2(M-\omega)}{\frac{dH}{\dot{r}}dr},
\end{equation}
for Hamiltonian $H = M-\omega^\prime$ and trajectory of emission is given by radial null geodesics  $\dot{r}=1-\sqrt{\frac{2(M-\omega^\prime)}{r}}$.  We remark that the emitted mass $\omega$ has been substrated from $M$ such that total energy is conserved.  Now we can apply the Sommerfeld-Bohr quantization rule and demand that each emission of $\omega$ carries away the action quantum $h$, which corresponds to a decrease of the horizon area.  That is,
\begin{equation}\label{WKB}
2 Im \int_{r_{in}}^{r_{out}}{p_r dr}  = h.
\end{equation}
This brings down to a simple quadratic equation of $\omega$:
\begin{equation}\label{quantization}
(2-\frac{\alpha\hbar}{2\pi M^2})\omega^2  -(4 M+\frac{\alpha\hbar}{\pi M})\omega + \hbar=0
\end{equation}
Were both nonthermal and quantum effect effects ignorable, one can drop the $\omega^2$ term and obtain the quantum of mass $\omega = \frac{\hbar}{4M}$ by solving (\ref{quantization}).  The area discreteness can be computed as
\begin{equation}
\Delta A = 8 \pi r \frac{dr}{dM} \Delta M \big|_{r=2M,\Delta M = \omega} = 8 \pi \hbar
\end{equation}
We remark that $\hbar=\l_p^2$ in our choice of unit. The universal prefactor $8\pi$ agrees with that obtained from previous methods\cite{Jiang:2012dm}.  Now we would like to include the nonthermal and quantum effects by solving (\ref{quantization}) honestly and obtain
\begin{equation}
\omega \simeq \frac{\hbar}{4 M} + \frac{\hbar^2(\pi-2\alpha)}{32\pi M^3} + {\cal O}(\frac{\alpha^2}{M^5}),
\end{equation}
where we choose the smaller root for $\omega < M$ and do the Taylor expansion as long as $M \gg \l_p$.  At last, we have the area discreteness
\begin{equation}\label{discreteness}
\Delta A = 8 \pi \hbar + \frac{(16\pi^2-32\pi\alpha ) \hbar^2}{A} + \cdots
\end{equation}
Due to the nonthermal correction, the area spacing gets larger as the horizon area shrinks as $\alpha<\frac{\pi}{2}$ but gets smaller vice versa.  This can be regarded as an important signature for the Parikh-Wilczek tunneling model of Hawking radiation if the area discreteness were ever to be detected in the future.  The area discreteness can be easily generalized to the Schwarzschild black hole in arbitrary dimension $D$\cite{Wei:2014bva}, where
\begin{equation}\label{discreteness_D}
\Delta A = 8 \pi \hbar^{D/2-1} + \frac{(32\pi^2-32\pi (D-2)\alpha) \hbar^{(D-2)}}{(D-2) A} + \cdots,
\end{equation}
where $A=r_0^{D-2}\Omega_{D-2}$ for horizon radius $r_0$.  We remark that in $D=4$ the nonthermal correction is competitable to the quantum correction, however the former becomes less and less important as $D$ increases.

To obtain correction to black holes with more charges or different topology, one can in principle solve the following algebraic equation as a generalization of (\ref{quantization}):
\begin{equation}\label{algebra_eqn}
S_{BH}(Q_i-q_i)-S_{BH}(Q_i) + 2\pi =0,
\end{equation} 
given the change of black hole entropy as a function of black hole charges $Q_i$ and emitted charges $q_i$.  In the following examples, we will solve (\ref{algebra_eqn}) in the large mass limit, i.e. $M \gg 0$, and focus on nonthermal correction but ignore the quantum correction:
\begin{itemize}
\item For a Reissner-Nordstr\"{o}m black hole of mass $M$ and electric charge $Q$, we have metric
\begin{eqnarray}
&&ds^2 = -f(r) dt^2 + f^{-1}(r) dr^2 + r^2 d\Omega_2^2,\nonumber\\
&&f(r) = 1-\frac{2M}{r} + \frac{Q}{r^2},
\end{eqnarray}
where the horizon $r_+ = M + \sqrt{M^2-Q^2}$.  It is convenient to define the extremality $\Gamma = Q/M$ and charge-mass-ratio of emitted particle $\gamma=q/\omega$.  The area discreteness in generic reads
\begin{equation}
\Delta A = 8 \pi \hbar (1 + a(\Gamma,\gamma))+ \frac{16\pi^2\hbar^2}{A} (1+b(\Gamma,\gamma)),
\end{equation}
where functions $a(\Gamma,\gamma)$ and $b(\Gamma,\gamma)$ are complicated but can be perturbatively computed.  For instance,
\begin{equation}
a(\Gamma,\gamma) \simeq \frac{1}{2}\gamma\Gamma + {\cal O}(\Gamma^2),\quad b(\Gamma,\gamma) \simeq -\frac{\gamma^2}{2} +\frac{3}{2}\gamma \Gamma -\frac{3}{4}\gamma^3\Gamma + {\cal O}(\Gamma^2),
\end{equation}
in the near Schwarzschild limit ($\Gamma \ll 1$).  On the other hand, in the near extremal limit where $\Gamma,\gamma \to 1$, we obtain
\begin{equation}
a(x,\gamma) \simeq 3-12x + {\cal O}(x^2),\quad b(x,\gamma) \simeq 3-22x + {\cal O}(x^2),
\end{equation}
where $\Gamma \equiv 1-2x^2$.

\item For a  BTZ black hole in three dimensisons\cite{Banados:1992wn,Martinez:1999qi}, area spectrum has been discussed in \cite{Kwon:2010um,Li:2013} and tunneling rate was discussed in \cite{Ejaz:2013fla}.    We begin with the metric
\begin{eqnarray}
&&ds^2 = -f(r) dt^2 + f^{-1}(r) dr^2 + r^2 (d\phi-\frac{J}{2r^2}dt)^2,\nonumber\\
&&f(r) = -M +\frac{r^2}{\l^2}+\frac{J^2}{4r^2}
\end{eqnarray}
The entropy function is known
\begin{equation}
S_{BH} = \frac{\pi}{2\hbar} r_+,
\end{equation}
where $r_+^2 = \frac{1}{2}\big(M\l^2+\sqrt{M^2\l^4-J^2\l^2}\big)$.  Following the (\ref{algebra_eqn}), one obtains for nonrotating BTZ ($J=0$):
\begin{equation}
\Delta A = 8\pi\hbar - \frac{32\pi^2\hbar^2}{A} + \cdots.
\end{equation}
For $J \neq 0$, the area spacing in generic depends on the black hole angular momentum $J$ and spin of the emitted particle $j$.  If one defines the extremality $\Gamma \equiv J/M$ and emitted particle's spin-mass-ratio $\gamma \equiv j/\omega$, then 

\begin{equation}\label{BTZ_space}
\Delta A = 8\pi \hbar - \frac{32\pi^2\hbar^2}{A}(1+ a(\Gamma,\gamma)) + \cdots,
\end{equation}
where functios $a(\Gamma,\gamma)$ can be solved by Taylor's expansion at small $\Gamma$ and $\gamma$:
\begin{equation}
a(\Gamma,\gamma) \simeq  \frac{\gamma^2}{\l^2} -2\frac{\gamma\Gamma}{\l^2} -\frac{\Gamma^2}{8\l^2} + \cdots.
\end{equation}
The constant leading term in (\ref{BTZ_space}) agrees with that found in \cite{Kwon:2010um,Li:2013}, and moreover we observe the nontermal correction depends on $\Gamma$ and $\gamma$ in general.

\item For a $D$-dimensional AdS black hole of different horizon topologies:
\begin{eqnarray}
&&ds^2 = -f(r) dt^2 + f^{-1}dr^2 + r^2 d\Omega_{D-2}^2,\nonumber\\
&&f(r) = k + \frac{r^2}{\l^2} - \frac{aM}{r^{D-3}}.
\end{eqnarray}
For simplicity, we first examine the one with planar horizon, that is $k=0$.  The horizon can be analytically solved as  $r_+=(a\l^2 M)^{\frac{1}{D-1}}$.   We find leading-order equisistant spectrum:
\begin{equation}
\Delta A = 8\pi \hbar^{D/2-1}- \frac{32\pi^2 \hbar^{(D-2)}}{(D-2)A} + \cdots.
\end{equation}
The leading term agrees with the universal factor $8\pi$ for any $D>3$, but  different from that obtained in the \cite{Daghigh:2008jz}.  The nonthermal correction takes same form as that in the Schwarzschild black hole (\ref{discreteness_D}) but with opposite sign.  We remark that the correction implicitly depends on the AdS radius of curvature $\l$ via the horizon area $A$.  This result cannot be simply compared with (\ref{discreteness_D}) in the flat limit $\l \to \infty$ due to different horizon topology $k=0$.

For the spherical near-horizon topology, $k=1$, we find that correction to area spectrum explicitly depends on $\l$.  In particular, at the limit of large mass and weak curvature (but keep $M/\l$ small), one obtains
\begin{equation}
\Delta A \simeq 8\pi\hbar +\frac{\pi\hbar^2}{M^2} + \l^{-2}(4M\hbar-\frac{\hbar^3}{8M^3}) +\cdots,
\end{equation}
for $D=4$.  We remark that the result of (\ref{discreteness_D}) can be reproduced in the flat limit $\l \to \infty$.

\item For a $D$-dimensional Schwarzschild-de Sitter black hole:
\begin{equation}
ds^2 = -(1-\frac{2M}{r^{D-3}}-\frac{r^2}{\l^2})dt^2+(1-\frac{2M}{r^{D-3}}-\frac{r^2}{\l^2})^{-1}dr^2+r^2d\Omega_{D-2}^2,
\end{equation}
First, we would like to examinate the case of $D=3$, where one obtains exact solution
\begin{equation}
\Delta A = 8\pi \hbar
\end{equation}
Since there is in fact no black hole in three dimensional de Sitter space, this should be identified as the area spectrum of $dS_3$ space itself\footnote{As discussed in \cite{LopezOrtega:2009ww}, if additional contribution due to volume change is included, we would obtained twice of discreteness as $\Delta A = 16\pi \hbar$.}.

For $D>3$, one recieves the area spectrum correction.  For instance, in $D=4$ for large $M$ and $\l$:
\begin{equation}\label{SSdS}
\Delta A \simeq 8\pi \hbar -  2592\pi \hbar 2^{2/3}(\frac{M}{\l})^{8/3} +\cdots
\end{equation}

\item For a $D$-dimensional AdS topological black hole\cite{Birmingham:1998nr}:
\begin{equation}
ds^2 = -(-1-\frac{2M}{r^{D-3}}+\frac{r^2}{\l^2})dt^2+(-1-\frac{2M}{r^{D-3}}+\frac{r^2}{\l^2})^{-1}dr^2+r^2d\Omega_{D-2}^2,
\end{equation}
We obtain $D=4$ spectrum for large $M$ and $\l$:
\begin{equation}
\Delta A \simeq 8\pi \hbar -  2592\pi \hbar 2^{2/3}(\frac{M}{\l})^{8/3} +\cdots,
\end{equation}
which is the same as (\ref{SSdS}).  For massless topological black hole, where $M\to 0$, we obtain the universal result $\Delta A = 8\pi\hbar$, which again is half of what was found in \cite{LopezOrtega:2010wx} where the volume change is included.

\item For a $D$-dimensional Gauss-Bonnet black hole, the metric reads:
\begin{eqnarray}
&&ds^2=-f(r)dt^2+f(r)^{-1}dr^2+r^2d\Omega_{D-2}^2,\nonumber\\
&&f(r) = 1+\frac{r^2}{2\alpha}\big[ 1- (1+\frac{4\alpha a M}{r^{D-1}})^{1/2} \big], \nonumber\\
&& \alpha = (D-3)(D-4)\alpha_{GB}, \quad a =\frac{16\pi G}{(D-2) \Omega_{D-2}}.
\end{eqnarray}
The tunneling model of the (AdS) Gauss-Bonnet black hole has been studied in \cite{Huang:2006,Muneyuki:2011jm}, and the emission rate agrees with that in (\ref{tunnel_rate}), where the entropy is given by 
\begin{equation}
S=\frac{r_+^{D-2}\Omega_{D-2}}{4}\big[ 1+2 (\frac{D-2}{D-4})\frac{\alpha}{r_+^2} \big],
\end{equation}
where $r_+$ satisfies 
\begin{equation}
 a M r_+^{5-D} =  r_+^2 + \alpha.
\end{equation}
The area spectrum was discussed in \cite{Ren:2010zz}, where the conclusion that entropy spectrum is shown to be equally spacing agrees with our assumption (\ref{algebra_eqn}).  In particular, the coefficient $\alpha$ vanishes for $D=4$ such that
\begin{equation}
\Delta A = 8 \pi \hbar+ \frac{16\pi^2\hbar^2}{A} + \cdots 
\end{equation}
which has no effect from the Gauss-Bonnet term.  For $D=5$, the area spectrum correction can be expressed via Taylor's expansion of $\alpha_{GB}/M$:
\begin{eqnarray}
&&\Delta A= 8 \pi \hbar( 1+a(\alpha_{GB},M) ) + \frac{32\pi^2\hbar^2}{3A} (1+b(\alpha_{GB},M))+\cdots ,\nonumber\\
&&a(\alpha_{GB},M)=-\frac{3\pi}{2M}\alpha_{GB}+\frac{9\pi^2}{8M^2}\alpha_{GB}^2 +\cdots,\nonumber\\
&&b(\alpha_{GB},M)=-\frac{3\pi}{2M}\alpha_{GB}+\frac{9\pi^2}{16M^2}\alpha_{GB}^2 +\cdots,
\end{eqnarray}

\end{itemize}

\begin{figure}[tbp]
\includegraphics[width=0.6\textwidth]{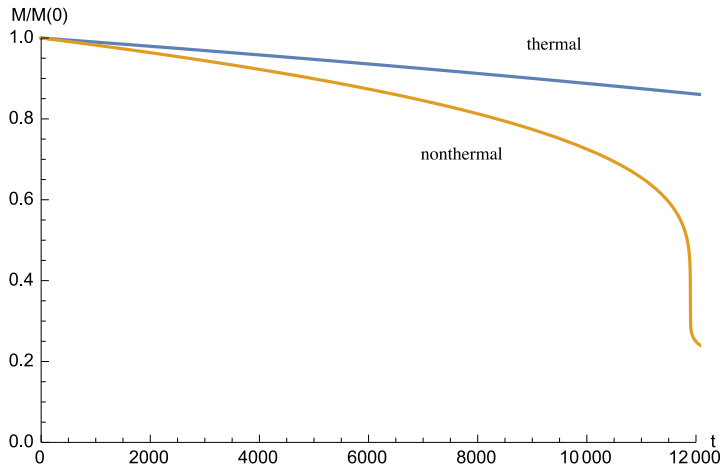} 
\caption{\label{fig:evolution} Time evolution for usual Hawking radiation as blackbody radiation, and radiation with nonthermal correction.  The latter evaporates faster due to increasing spacing of area spectrum. }.
\end{figure}

In summary, we have investigated the nonthermal correction to area spectrum in various kinds of black holes using the quantization rule (\ref{WKB}).   This semiclassical approximation usually works better for highly excited states, that is, large black hole mass (charges), and the leading term reproduces the universal coefficient $8\pi$.  However, if the equidistant spectrum for entropy, $\Delta S_{BH}=2\pi$, could persist through the lifetime  of a black hole,  (\ref{WKB}) predicts an increasing correction to area spectrum toward the end of evaporation.  To estimate the nonthermal correction to the lifetime of a Schwarzschild black holes, 
we observe in (\ref{discreteness}) that nonthermal correction contributes like a quantum correction with $\alpha = -\frac{\pi}{2}$ at the ${\cal O}(1/A)$ order.  Therefore, the nonthermal effect could be modeled as radiation at an {\sl effective} tmeperature 
\begin{equation}
T^{eff}_{H}=\frac{1}{8\pi M}(1-\frac{1}{8M^2})^{-1}.
\end{equation}
In the figure \ref{fig:evolution}, we plot both thermal radiation and nonthermal radiation for the Schwarzschild black hole.  It is expected that black hole speeds up its evaporation in the nonthermal radiation thanks to increasing spacing in area spectrum. 

\begin{acknowledgments}
We are grateful to useful discussion with Pisin Chen, Feng-Li Lin, Hsien-Chung Kao, Otto Kong and Cheng-Wei Chiang.  We thank June-Yu Wei and Mei-Hsian Wang for their participation in early stage.  This work is supported in part by the Taiwan's Ministry of Science and Technology  (grant No. 102-2112-M-033-003-MY4) and the National Center for Theoretical Science. 
\end{acknowledgments}


\end{document}